\documentclass[letter]{aa} 
\usepackage[varg]{txfonts}
\usepackage{graphicx}
\usepackage{txfonts}
\usepackage{multirow}
\usepackage{booktabs,caption}
\usepackage[flushleft]{threeparttable}
\usepackage{graphicx}
\usepackage{pdflscape}
\usepackage{amsmath,amssymb}
\usepackage{float,lscape}
\usepackage{longtable}
\usepackage{xcolor}
\usepackage{tabularx}
\usepackage{array}
\usepackage{rotating}
\usepackage{makecell}
\usepackage{tablefootnote}
\usepackage{booktabs}
\usepackage{threeparttable}
\usepackage{siunitx}
\usepackage{soul}
\usepackage[utf8]{inputenc}
\usepackage{hyperref}
\DeclareUnicodeCharacter{02BC}{'}

\begin{document}

   \title{Star-formation in neutral hydrogen gas reservoirs at cosmic noon}


   \author{Dharmender
          \inst{1},
          Ravi Joshi\inst{2},
          Michele Fumagalli\inst{3,4},
          Pasquier Noterdaeme\inst{5},
          Hum Chand\inst{1}
          \and
          Luis C. Ho\inst{6,7}
          }

   \institute{Central University of Himachal Pradesh,
              Dharamshala,Kangra, India\\
              \email{dharmender98thakur@gmail.com}
         \and
             Indian Institute of Astrophysics (IIA), Bengaluru, Karnataka, India\\
             \email{rvjoshirv@gmail.com}
               \and
          Universit\'a degli Studi di Milano-Bicocca, Dip. di Fisica G. Occhialini, Piazza della Scienza 3, 20126 Milano, Italy
         \and 
            INAF—Osservatorio Astronomico di Trieste, via G.B. Tiepolo 11, I-34143 Trieste, Italy
          \and
          Institut d'Astrophysique de Paris, CNRS-SU, UMR 7095, 98bis bd Arago, 75014 Paris, France
        \and
        Kavli Institute for Astronomy and Astrophysics, Peking University, Beijing 100871, Peopleʼs Republic of China
             \and
        Department of Astronomy, School of Physics, Peking University, Beijing 100871, Peopleʼs Republic of China
             }

  \titlerunning{star formation in DLAs}
  \authorrunning{Dharmender et al.}
  \abstract
  {We aim to constrain the average star formation associated with neutral hydrogen gas reservoirs at cosmic noon. Using a unprecedented sample of 1716 
high column density Damped Ly-$\alpha$ absorbers (DLAs) from the Sloan Digital Sky Survey with log($N$(H{~\sc i}) / \rm cm$^{-2}$) $\ge$21, we generated the average Ly-$\alpha$ emission spectrum associated to DLAs, free from emission from the background quasar. We measured Ly$\alpha$ emission at $> 5.8\sigma$ level with luminosity $8.95\pm 1.54 \times \rm  10^{40}\ \text{erg}\ \text{s}^{-1}$ (corresponding to about 0.02 L$^{\star}$ at $z \sim$ 2-3) in systems with average log($N$(H{~\sc i}) / \rm $cm^{-2}$) $\approx$21.2 and at median redshift of $z \sim$ 2.64.
The peak of the Ly$\alpha$ emission is apparently redshifted by $\sim$300~km\,s$^{-1}$  relative to the absorption redshift, which is seemingly due to suppression of blue Ly-$\alpha$ photons by radiative transfer through expanding gas.
We infer that DLAs form stars with an average rate of (0.08 $\pm$ 0.01)/$\text{f}_\text{esc}\ \rm \text{M}_{\odot}\ \text{yr}^{-1}$, i.e, $\approx (0.54\pm 0.09)\rm \text{M}_{\odot}\ \text{yr}^{-1}$ for a typical escape fraction, $\text{f}_{\text{esc}} =0.15$, of Lyman-$\alpha$ emitting galaxies. DLA galaxies follows the star formation main sequence of star-forming galaxies at high redshift, suggesting that the DLA population is dominated by the lower mass end of Lyman-$\alpha$ emitting galaxies.}
   \maketitle

\section{Introduction}

The formation and evolution of galaxies are driven by the filamentary accretion of pristine gas from the cosmic web, which feeds the interstellar medium (ISM) and fuels star formation \citep{Barnes_2009,fumagalli2011absorption,bird2014damped,faucher2023key}. Being the most abundant element in the Universe, hydrogen is a unique tracer that probes gas across various scales, ranging from interstellar to circumgalactic and intergalactic medium. However, mapping H{\sc~i} gas in galaxies emission is difficult even at moderate redshifts \citep{Bera2018ApJ...865...39B}. At high redshift, $z \gtrsim 2$, the Damped Ly$\alpha$ (DLA) absorbers, characterized by hydrogen column densities of \( \log(N_{\text{H {\sc~i}}} / \text{cm}^{-2}) \geq 20.3 \), offer a direct probe of the distribution of neutral gas at a mean  density around one-tenth of the star formation threshold \citep{Cen2012ApJ...748..121C}. DLAs contain a large fraction of the neutral hydrogen in the Universe and are believed to be the progenitors of normal disk galaxies \citep{wolfe2005damped, Neeleman2020Natur.581..269N,kaur2024hi}. They are associated with star-forming regions, as evidenced by their enrichment with redshift \citep{rafelski2012metallicity,neeleman2018molecular}.

To date, several thousand DLAs at $z > 2$ have been discovered in the Sloan Digital Sky Survey \citep{prochaska2004sloan, noterdaeme2009evolution, noterdaeme2012discovery,chabanier_completed_2022-1}, and hundreds of these have been followed for high-resolution spectroscopy \citep{ho2020detecting}. Despite these advances, our understanding of the origin of DLAs and the physical characteristics of the associated galaxies remains limited. Earlier searches based on long-slit spectroscopic measurements yielded a low detection rate of $\sim$10 percent which is attributed to factors such as bright background quasars, dust attenuation, or the possibility that only a subset of the DLA population is directly associated with active star formation. The efforts to map the metal-rich DLAs resulted in a higher detection rate at $\sim$64 percent \citep{krogager2017consensus}, but provide limited information on star formation rates, masses, extent of H{\sc~i}, and large-scale environments \citep{krogager2017consensus}. Despite the numerous efforts to search for DLA host galaxies in emission, until recently, only $\sim$25 DLA host galaxies have been discovered so far \citep{christensen_2014,krogager2017consensus}.  \par

More recently, using the large 3D field of MUSE/VLT \citet{Fumagalli2017MNRAS.471.3686F}  have detected a tantalizing example of a diffuse gas environment of about 50 kpc near a $z \sim 3.2$ DLA. \citet{Mackenzie2019MNRAS.487.5070M} have searched  six DLAs using  MUSE/VLT with $z \ge 3 $ within 1000 $\rm km\ s^{-1}$ and found a high detection rate of $\sim$80 percent with impact parameters between 25 and 280 kpc \citep[see also,][]{lofthouse2023muse}. Furthermore,  interferometers such as the ALMA have overcome the dust bias, with the detection of a few tens of molecular gas-rich systems using CO rotational transitions and the atomic [C{\sc~ii}] line 
\citep{neeleman2013fundamental,neeleman2017c,neeleman2019c,lofthouse2023muse,kaur2024hi}. So far, these studies have focused on tracing relatively high-metallicity systems, finding the DLA host at relatively large impact parameters, $\sim16-45$ kpc, and high molecular gas masses of $10^{10}-10^{11}\text{M}_{\odot}$. These efforts have revealed that DLA host galaxies are similar to massive star-forming galaxies and are embedded in enriched neutral hydrogen gas reservoirs that extend well beyond the star-forming interstellar medium of these galaxies. \par

In efforts to study the average properties of DLAs, \citet{fumagalli2014directly} measured the rest-frame far-ultraviolet flux from DLAs in HST composite images using the `double-DLA' technique, that uses two optically thick absorbers to eliminate the glare of the bright background quasars \citep[see also,][]{Omeara2006ApJ...642L...9O,Christensen2009A&A...505.1007C}. They have excluded the presence of compact star-forming regions with SFRs as low as $\sim0.1 \rm M_{\odot}\ yr^{-1}$ at the position of the quasar. On the other hand, the spectral stacking experiments to detect the Ly$\alpha$ emission from the DLA host galaxy have mostly set an upper limit on Ly$\alpha$ luminosity and SFR    \citep{rahmani2010lyman, Cai2014ApJ...793..139C,joshi2016lyalpha}, except for metal-rich DLAs \citep{joshi2016lyalpha}.  Interestingly, \citet{Noterdaeme2014A&A...566A..24N} have shown that the extremely strong DLAs resemble the Lyman-$\alpha$ emitting galaxies (LAEs) with Ly$\alpha$ luminosity  of $\ge 0.7 \times 10^{42} \rm erg\ s^{-1}$.
In this letter, we examine DLA systems with high neutral hydrogen column densities, log($N_{\text{H {\sc~i}}} / \text{cm}^{-2}) \ge$ 21, which is five times lower than the threshold used by \citet{Noterdaeme2014A&A...566A..24N}. Using a dataset twice the size of \citet{joshi2016lyalpha}, we present a clear detection of Ly$\alpha$ emission in these systems and explore the nature of the DLA host galaxies. The sample selection and analysis are provided in Section~\ref{sample}, and Section~\ref{analysis}. The obtained results are given in Section~\ref{results}, followed by a discussion and conclusion in Section~\ref{discussion}. Throughout, we have assumed the flat Universe with $\text{H}_0$ = 70 $\rm km\ s^{-1}\ Mpc^{-1}$, $\Omega_m$ = 0.3 and $\Omega_\Lambda$= 0.7.

\section{Sample Selection}
\label{sample}
Here, we utilize the SDSS-III BOSS DLA catalog from \citet{noterdaeme2012discovery} updated to Data Release-12 , consisting  19,535 DLA candidates with column density log($N_{\text{H {\sc~i}}} / \text{cm}^{-2}) \ge$ 20.3. 
In addition, we use  SDSS-IV Data Release-16 DLA catalog from \citet{chabanier_completed_2022} which is constructed based on a convolutional neural network,  comprising 30,019  DLAs with log($N_{\text{H {\sc~i}}} / \text{cm}^{-2}) \ge$ 20.3  and high confidence level of $\ge 0.5$.  To probe the average SFR associated with DLAs at the cosmic noon, the applied selection filters were (1) the log($N_{\text{H {\sc~i}}} / \text{cm}^{-2}) \ge$ 21 which ensures a dark core ($\tau >10$) of  Ly$\alpha$ absorption trough spread over at least 7 times
the average full width at half-maximum, FWHM ($\sim$160 $\rm km\ s^{-1}$) of the instrumental profile of the BOSS spectrograph; (2) we avoid the proximate DLAs by considering the systems with velocity offset of $\ge$ 5000 \rm km\ s$^{-1}$ relative to the quasar emission redshift and exclude the sightlines with broad absorption lines from quasar outflows;  (3) to ensure an accurate determination of the H{\sc~i} column density,  we impose a continuum to noise ratio (CNR) $\ge$ 4 in the spectral region of $\sim$20$\AA$ blue and redder side of the DLA wing; (4) to avoid the confusion with Ly$\beta$ forest we only consider the DLAs within in Ly$\alpha$ forest region. The aforementioned selection criteria resulted in 1716 absorbers over a redshift range of $1.97 < z < 4.0$ with a median $ z = 2.637$ and a neutral hydrogen column density ranging between $ 21.0 \le \log(N_{\text{H {\sc~i}}} / \text{cm}^{-2}) \le 22.3$, with median $\log(N_{\text{H {\sc~i}}} / \text{cm}^{-2}$)= 21.2. 
The sample summary is given in Table~\ref{tabsample}.

Finally, to avoid any uncertainty in the absorption redshift and/or H{\sc~i} column density measurements, we further visually inspected the entire sample, and measured the absorption redshift by cross-correlating the low-ionization metal absorption lines (e.g. \text{CII} $\lambda 1334, \text{SiII} \lambda 1526, \text{AlII} \lambda 1670 $, $\text{FeII} \lambda\lambda 1608, 2344, 2374, 2382, 2586, 2600, \text{MgII} \lambda\lambda 2796, 2803$) redward to the quasar Ly$\alpha$ emission. For the systems without a metal absorption line detection, we keep the redshift obtained from the H{\sc~i} Lyman series transitions based on the CNN  models \citep{chabanier_completed_2022}. This resulted in a final sample of 1716 (excluding duplicates among DR12 and DR16) unique DLAs,  of which 1515 systems have redshift determined from metal lines.
\section{Analysis}
\label{analysis}
To detect faint Ly$\alpha$ emission, we generate the stacked spectra by shifting the individual spectrum to the rest-frame of the DLA, while conserving the flux and re-binned to a common grid, keeping the same pixel size (constant in velocity space) as the original data. To remove the contribution from any possible outliers, we have used the median and $3\sigma$-clipped weighted mean statistics with 1/$\sigma^2$ weighting, with $\sigma$ being the error in the flux \citep[see also,][]{joshi2016lyalpha}.  Considering the wide flat, dark absorption trough of DLAs, we do not re-scale or normalize the spectrum before the co-addition. Further, to test the robustness we generate  5000 bootstrapped stacked spectra and estimate the $1\sigma$ uncertainty over each pixel as 16th and 84th percentiles of the flux distribution of the corresponding pixel. \par

Note that the residual flux level within the core of the DLA absorption profile exhibits a non-zero offset, likely related to sky-subtraction errors \citep{rahmani2010lyman, joshi2016lyalpha} and/or average FUV radiation from the background quasar host galaxies \citep{Cai2014ApJ...793..139C}. Our stacked DLA profile also shows a positive flux with no prominent signal blue-ward of Ly$\alpha$ ( see, Figure\ref{stack_nonzero} in the Appendix A). We correct this offset by subtracting the weighted mean flux in the dark trough of DLA, estimated using the pixels with $\tau > 10$, blue-ward of the line center at 1215.67\AA\ where the IGM absorption may likely result in zero flux from DLA galaxy. This correction may potentially eliminate any continuum emission from the DLA host, but our primary interest here is on the Ly$\alpha$ emission line.

\begin{table}[]
\caption{DLA sample selection }
\label{tabsample}
\begin{tabular}{lll}
\hline
Criteria & SDSS DR12  & SDSS DR16  \\ 
\hline 
log($N_{\text{H {\sc~i}}} / \text{cm}^{-2}) \ge$ 20.3 & 19,535 & 30,019 (Conf $>$ 0.5) \\
1.97 $\le$ z$_{abs} \le$ 4 &  \\  
log($N_{\text{H {\sc~i}}} / \text{cm}^{-2}) \ge$ 21.0  & 1139 & 1531  \\
$\beta \ge$ 5000 km s$^{-1}$ &  \\ 
non-BAL & \\
z$_{abs}$ > z(Ly$\beta$) & &\\ 
CNR $\ge$ 4 & 1139 & 930 \\ 
unique DLAs in DR16 &-- & 577 \\
\hline 
Total Sample & & 1716 \\ 
\hline
\end{tabular}
\end{table}
\begin{figure*}
    \centering
    \includegraphics[width=0.85\textwidth,height=0.33\textwidth]{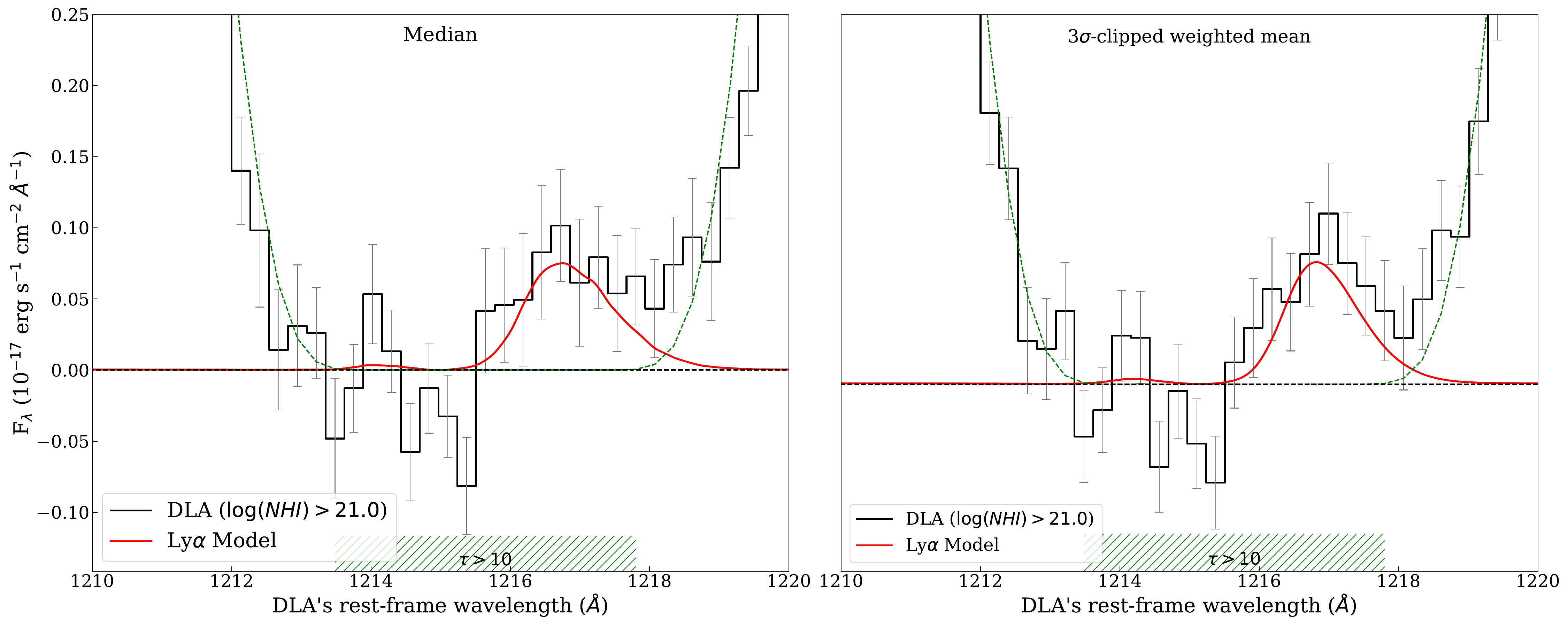}
    \caption{The median ({\it left panel}) and 3$\sigma$-clipped weighted mean ({\it right panel}) stacked spectrum for 1716 DLAs with median hydrogen column density of log($N_{\text{H {\sc~i}}} / \text{cm}^{-2}$)=21.2, along with 1$\sigma$ error shown in grey color. The green hatched region show the dark core of absorption trough, with optical depth $(\tau) > 10$. The red curve shows the emergent Ly$\alpha$ profile, after radiative transfer of Ly$\alpha$ photons while considering the DLAs as expanding Thin-Shell geometry (see text, \citet{gurung2022zelda}).}
    \label{fig:1}
\end{figure*}
\section{Results}
\label{results}
In Figure~\ref{fig:1}, we show the median and  $3\sigma$-clipped  weighted mean stacked spectra of Ly$\alpha$ absorption trough. 
We detect an asymmetric flux distribution, with enhanced flux redward of the line center, i.e. 1215.67\AA\  in the dark core of the absorption trough. Such Ly$\alpha$ emission profiles with enhanced red and suppressed blue peaks are naturally produced by the radiative transport of Ly$\alpha$  photons in an outflowing medium 
\citep{steidel2010structure,Dijkstra2014PASA...31...40D,galbiati2023muse}, also observed empirically \citep{steidel2010structure, Galbiati2023MNRAS.524.3474G}. To quantify the emission line flux, at first, we use the region redward of the line center and detect a strong Ly$\alpha$ emission profile at $5.8\sigma$ level with Ly$\alpha$ flux of $(0.150 \pm 0.026) \times  10^{-17}  \rm erg \ s^{-1}\ cm^{-2}$ in the 3$\sigma$-clipped weighted-mean stack. Similarly, for the median stack the line is detected at 4.6$\sigma$ level with Ly$\alpha$ flux of $(0.145 \pm 0.032) \times  10^{-17}  \rm erg \ s^{-1}\ cm^{-2}$.  At the median redshift of our DLA sample, i.e., $z \sim 2.64$ this corresponds to a Ly$\alpha$ luminosity of $(8.64 \pm 1.87) \times 10^{40} \rm erg\ s^{-1}$ in the median and  $(8.95 \pm 1.54) \times 10^{40} \rm erg\ s^{-1}$ in the 3$\sigma$-clipped weighted mean stack. We get a similar luminosity of $(6.45 \pm 2.50) \times 10^{40} \rm erg\ s^{-1}$ at $2.57\sigma$ level and   $(6.68 \pm 2.13) \times 10^{40} \rm erg\ s^{-1}$ at $3.1\sigma$ level for median and 3$\sigma$-clipped weighted mean stacks, respectively, if we integrate all the pixels in the entire dark trough with $\tau > 10$. \par

Next, using the Ly$\alpha$ luminosity, we estimate the average SFR associated with DLA hosts, assuming a Ly\(\alpha\) to \text{H}\(\alpha\) ratio (Case B) recombination value of 8.7 \citep{kennicutt1998star}.  However, this provides the minimal SFR as dust obscuration and escape fraction may lower the observed Ly$\alpha$ emission compared to the intrinsic emission. The $\text{H}\alpha$ and SFR calibration [$\text{SFR} \, (\text{M}_\odot \, \text{yr}^{-1}) = 7.9 \times 10^{-42} \times \text{L}_{\text{H}\alpha} \, (\text{erg} \, \text{s}^{-1})$] from \citet{kennicutt1998star} results in an average star formation rate $\langle \text{SFR} \rangle \sim 0.078 \pm 0.016 \, \text{M}_\odot \, \text{yr}^{-1}$. Note that the above estimate is a lower limit of SFR as it does not take into account dust and consider the Ly$\alpha$ escape fraction ($\text{f}_\text{esc}$) of unity. Additionally, uncertainty in the DLA redshift could contribute to the emission line broadening, potentially leading to a loss of flux near the wings. 
Considering a typical escape fraction of $\text{f}_\text{esc} = $0.15 for high-$z$ LAEs with a typical Ly$\alpha$ emission line equivalent width of $< 20$\AA\ \citep{matthee2016production,goovaerts2024galaxy}, the average SFR in DLAs can be higher with $\langle \text{SFR} \rangle \approx  0.9 \text{L}_{\text{Ly}{\alpha}}/ \text{f}_\text{esc} \lesssim 0.52 \pm 0.11 \rm (\text{M}_{\odot}\ \text{yr}^{-1}$). Using the star formation rate surface density and mean gas surface density relation: $\rm \Sigma_{\text{SFR}}  (\text{M}_{\odot}\ \text{yr}^{-1}\ \text{kpc}^{-2}) = (2.5 \pm 0.7) \times 10^{-4}  \left( \frac{\Sigma_{gas}}{1 \text{M}_{\odot}\ pc^{-2}} \right)^{1.4 \pm 0.15} $ at low redshift from  \citet[][see their equation 4]{Kennicuttb1998ApJ...498..541K}, the lower column densities that are typical for DLAs log($N_{\text{H {\sc~i}}} / \text{cm}^{-2}) \approx \rm 21.2\ cm^{-2}$ in our sample gives an average $\Sigma_{\text{SFR}} = 8.75 \times 10^{-3} \rm (\text{M}_{\odot}\ yr^{-1}\ kpc^{-2})$. We note that only a small fraction, i.e., $\sim$15\% of DLAs exhibit H$_2$ gas; however, the correlation between SFR and H{\sc~i} in low-density gas in conjunction with the large fraction of our DLAs exhibiting metals hints at some level of $in-situ$ SFR \citep{balashev2018constraining}. A caveat in linking the inferred SFR with the DLA column density is that the measured H{\sc~i} density is not averaged over the disk, but a local estimate along a pencil beam. Local measurements may thus introduce more substantial scatter in the relation. A partial mitigation of this effect arises from considering the average emission in a large sample of DLAs, for which we consider a range of N(H{\sc~i}) values. Such an average estimate will bring us closer to the underlying values \citep{Barbieri2005A&A...439..947B}.
Following \citet[][see their equation 10]{Noterdaeme2014A&A...566A..24N}, using the above integrated SFR, one can obtain the DLA radius as $\rm \Sigma_{\text{SFR}} \pi r^2_{gal}\ =\ 0.078/ \text{f}_\text{esc}\ \text{M}_{\odot}\ \text{yr}^{-1}$, which results in an average galaxy size of $\sim$ 4 kpc and $\sim$7 kpc for a typical $\text{f}_\text{esc}$ of 0.15 and 0.05 for LAEs and LBGs respectively \citep{hayes2010escape,matthee2016production}.  Note that, for the DLA galaxies of the above sizes, due to the finite SDSS(BOSS) fiber size of the radius of 1  arcsec (1.5 arcsec) which corresponds to a physical size of 8(12) kpc at the median redshift of $z =$ 2.637, the measured fluxes suffer from the fiber losses as only a part of the galaxy may lie inside the fiber \citep{Lopez2012MNRAS.419.3553L}, thus the above measured SFR represents a lower limit on average SFR in DLAs.  \par
We find that the Ly$\alpha$ emission is predominantly red-shifted relative to the systemic velocity of the DLAs with a typical line offset of 
310 $\pm$ 52 \rm km\ s$^{-1}$ in the median stack. A similar velocity offset of 323 $\pm$ 36 \rm km\ s$^{-1}$ is found when considering the 3$\sigma$-clipped weighted mean stack. This velocity offset falls close to the typical peak shift of $\sim$230 km s$^{-1}$ measured in the case of LAEs  \citep{erb2014lyalpha,lofthouse2023muse} and LBGs which show a velocity offset of much less than 400 km s$^{-1}$ \citep[see,][]{Noterdaeme2014A&A...566A..24N,mackenzie2019linking}. The median (3$\sigma$-clipped weighted mean) stacked  Ly$\alpha$ emission line FWHM in DLAs is found to be $\sim$ 400$\pm$137\ (360$\pm$94) \rm km\ s$^{-1}$ which is slightly higher than the mean FWHM of the red peak of Ly$\alpha$ line in LAEs i.e., $\sim$260 km s$^{-1}$, but consistent with the LBGs FWHM of 364 km s$^{-1}$ \citep{trainor2015spectroscopic}. The excess line broadening can be reconciled with the redshift uncertainty of DLAs and the range of velocity offset and Ly$\alpha$ profiles from the DLA population, which introduces line broadening while stacking.

\begin{figure}
    \centering
\includegraphics[width=0.49\textwidth,height=0.42\textwidth]{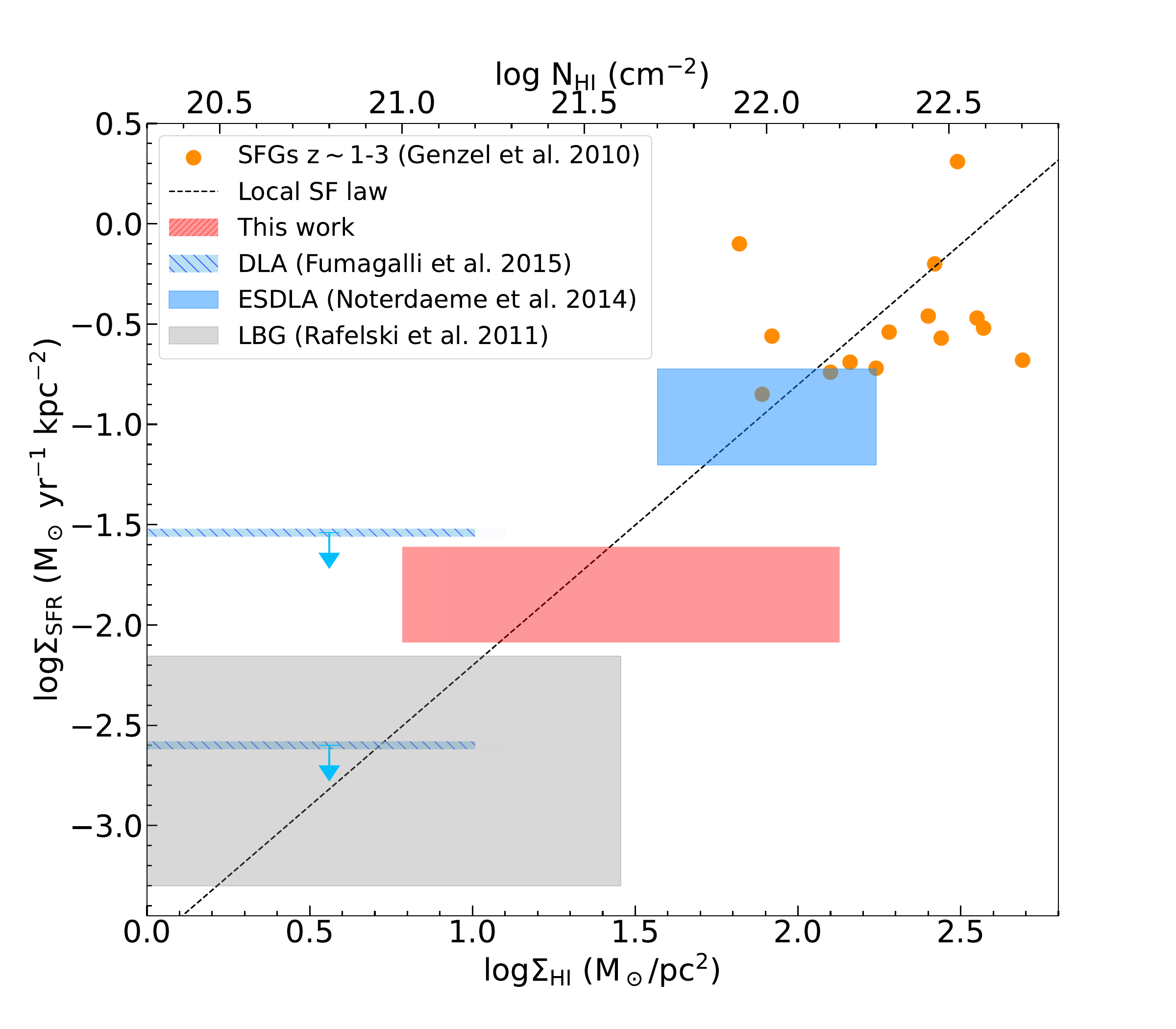}
    \caption{ The SFR surface densities $\rm \Sigma_{SFR}$ of DLAs from the spectral stacks. Limits on the SFR surface densities derived from the composite images are shown as horizontal hatched blue bars \citep{fumagalli2015directly}.  The grey shaded region show the $\rm \Sigma_{SFR}$ 
detected in the outskirts of LBGs  \citet{rafelski2011star}, and the blue shaded region shows the ESDLA ($N$(H{~\sc i}) > 10$^{21.7}$ \rm cm$^{-2}$) \citep{Noterdaeme2014A&A...566A..24N}. 
The dotted black line shows where the extrapolation of the local SF law. 
Our results shows log$N$(H{~\sc i}/cm$^{-2}) \ge$ 21 to 22.35 with range of f$_{\text{esc}}$, 0.05 and 0.15 showing dashed (red dashed) region.}
    \label{kslaw}
\end{figure}

\section{Discussion and Conclusions}
\label{discussion}
We use the largest set of 1716 high $N$(H{~\sc i}) DLAs to obtain direct limits on the {\it in-situ} SFRs in DLAs. Several spectral stacking experiments have been performed to detect the residual Ly$\alpha$ emission in the dark core of the DLA trough. \citet{rahmani2010lyman} studied the stacked spectra of DLAs with log($N_{\text{H {\sc~i}}} / \text{cm}^{-2}) \ge$ 20.62 and set  a 2$\sigma$ upper limit on the Ly$\alpha$ luminosity from DLA gas at $ \le 2 \times  10^{41}\ \rm erg\ s^{-1}$, which corresponds to a SFR of  $<  0.8 \rm \text{M}_{\odot}\ \text{yr}^{-1}$ at  $z\sim 2.8$. In another stacking experiment, \citet{Cai2014ApJ...793..139C} have used $\sim$2000 DLA systems with log($N_{\text{H {\sc~i}}} / \text{cm}^{-2})$> 20.6 at a median absorption redshift $\langle z \rangle$ = 2.6, and not detected Ly$\alpha$ emission signature in the dark trough of composite spectra.  Using $\sim$100 extremely strong DLAs with log($N_{\text{H {\sc~i}}} / \text{cm}^{-2}) $> 21.7,  which probe a small subset of DLA population,  \citet{Noterdaeme2014A&A...566A..24N} have detected the Ly$\alpha$ emission with luminosity of  $\sim 0.6 \pm 0.2 \times 10^{42} \rm \text{erg}\ \text{s}^{-1}$ (corresponding to about $0.1 L^{\star}$ at $z \sim 2 \ \rm to\ 3$). For DLA population of log($N_{\text{H {\sc~i}}} / \text{cm}^{-2})> 21$, \citet{joshi2016lyalpha} have detected a hint of Ly$\alpha$ emission at $2.8\sigma$ level in the red part of the DLA trough with luminosity  $< 10^{41}(3\sigma) \rm \text{erg}\ \text{s}^{-1}$. They found that the metal-rich systems show stronger Ly$\alpha$ emission. 
Here we have detected a strong signature of Ly$\alpha$ emission at $> 5\sigma$ level with an average luminosity of  $(8.64 \pm 1.87) × 10^{40}\ \rm erg\ s^{-1}$ and SFR of $\sim 0.078 \pm 0.016 \text{M}_\odot \, \text{yr}^{-1}$. This value is consistent with the average SFR detection limit (at 2$\sigma$) of 0.09 M$_\odot \, \text{yr}^{-1}$ observed over an aperture of 2kpc in direct HST imaging of DLA host galaxies with log($N_{\text{H {\sc~i}}} / \rm \text{cm}^{-2}$) of 20.3 to 21.3 by \citep{fumagalli2015directly}. 

\begin{figure}
    \centering
\includegraphics[width=0.46\textwidth,height=0.37\textwidth]{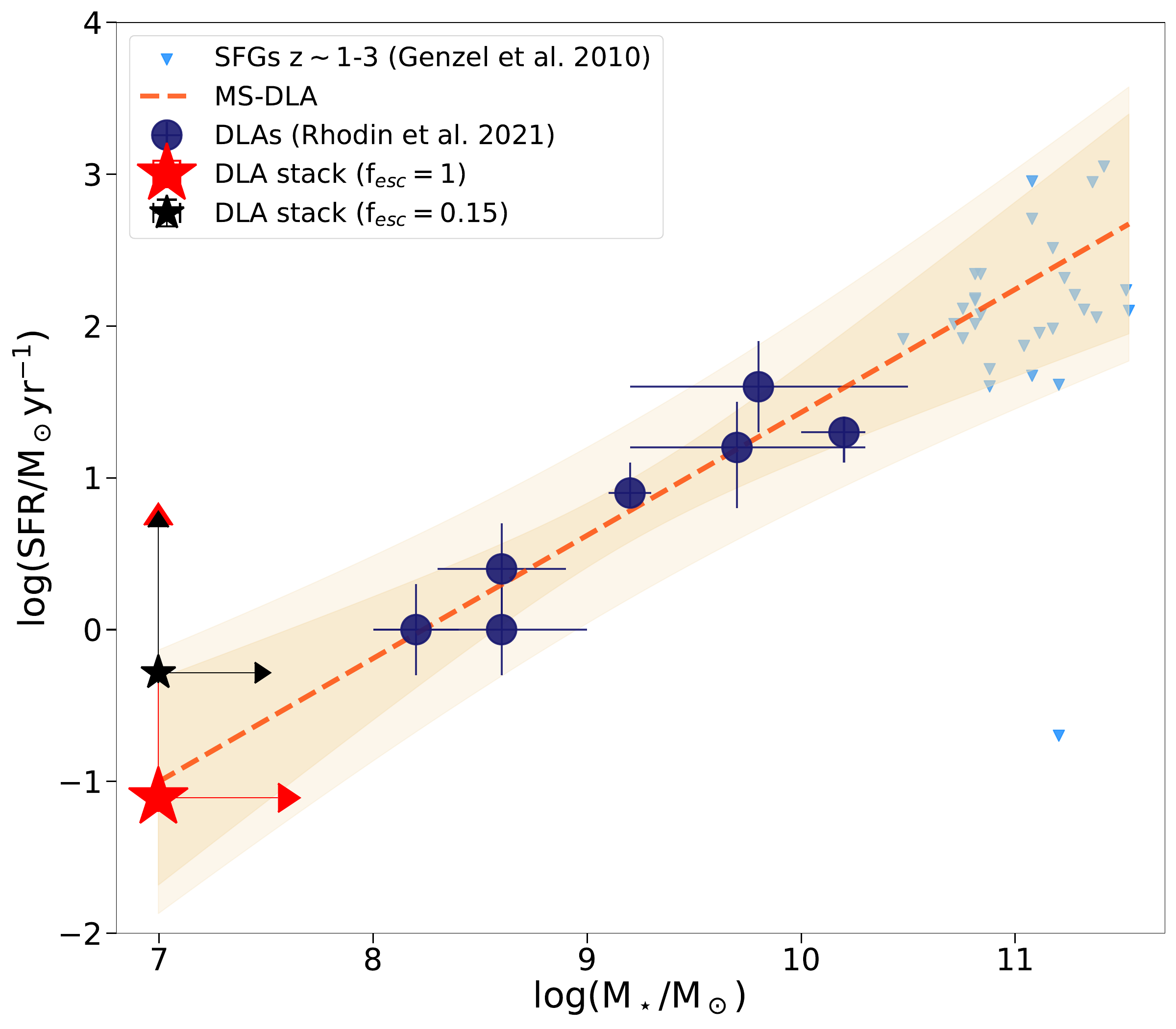}
    \caption{The star formation main sequence of star-forming galaxies ({\it triangle}) from \citet{genzel2010study}, and the direct detection of DLAs ({\it circle}) along with the best-fit MS relation ({\it dashed black line}) from \citet{rhodin2021absorption}. The stacked spectra measurements are shown as ({\it star}), whereas {\it arrow} represents the lower limit. 
    } 
    \label{smdla}
\end{figure}

We further explore the SFR relation with the observed neutral gas in absorption. Using our $in-situ$ SFR measurement we derive the star-formation surface density using a typical Lyman-alpha (Ly$\alpha$) halo scale length of approximately 4.5 kpc, derived from observations of high-redshift star-forming galaxies (3 $\leq z \leq$ 6) in the MUSE Hubble Ultra Deep Field Survey \citep{leclercq2017muse}. For such scale lengths the SDSS fiber probing a physical size of $\sim$12kpc likely integrates light from the DLA host.
In Figure~\ref{kslaw}, we plot the star-formation surface density ($\rm \Sigma_{SFR}$), assuming $f_{esc}$ = 0.05 and $f_{esc}$ =0.15 for LBGs and LAEs, with H{\sc i} gas surface density $\rm \Sigma_{gas}$. It is evident that the DLAs and ESDLAs follow the KS-law for a range of H{\sc~i} densities \citep[see also][]{Noterdaeme2014A&A...566A..24N, fumagalli2015directly}. \\
Previous efforts to study the DLA host galaxies suggest a strong association with LAEs \citep{Noterdaeme2014A&A...566A..24N,rubin2015dissecting,lofthouse2023muse}. By leveraging the average Ly$\alpha$ luminosity of DLAs, and incorporating the main sequence relation, we can estimate the stellar mass of DLAs using the equation presented by \citet{goovaerts2023evolution}, $\log$(\text{M}$_\star$ /\text{M}$_\odot$ ) = (0.85 $\pm$ 0.17) $\log$(L$_{\text{Ly}\alpha}$ / erg s$^{-1}$) $-$ (27.8 $\pm$ 6.4), yielding an average stellar mass $\log$(\text{M}$_\star$/\text{M}$_\odot)$ = 6.9 $\pm$ 0.6.
In Figure \ref{smdla}, we compare the stellar mass and average SFR in DLAs with SFGs at $z \sim$ 1 to 3 \citep{genzel2010study}, and a handful of known DLA host galaxies \citep{rhodin2021absorption}. It suggests that on an average the DLAs with log($N_{\text{H {\sc~i}}} / \rm \text{cm}^{-2}$) $\ge 21$ probe the lower mass end of SFG.

We also find that the Ly$\alpha$ emission shows an asymmetric emission feature with a dominant red peak and suppressed blue peak with systemic velocity offset with respect to DLA profile of $\sim$310 $\pm$  52 km s$^{-1}$ (323 $\pm$ 36 km s$^{-1}$) in the median (3$\sigma$-clipped weighted mean) stack. This velocity shift implies a kinematic difference between the emitting and absorbing neutral hydrogen gas, potentially due to gas in-outflows or complex dynamics within these galaxies. Another possibility is that the DLA gas is rotating, with non-uniform star formation throughout \citep{ogura2020alma}. In order to obtain further insight into the gas geometry and kinematics in DLAs, we model the observed Ly$\alpha$ emission profile using the Thin Shell model, which consists of a central Ly$\alpha$ source, surrounded by an expanding outflowing thin, spherical shell of hydrogen and dust. It successfully reproduces the diversity of line profiles such as broad absorption in some Lyman break galaxies to asymmetric emission in LAEs \citep{Verhamme2006A&A...460..397V}. For this, we use the {\sc zELDA}\protect\footnote{zELDA \protect\url{https://github.com/sidgurun/Lya_zelda}} routine which uses an isothermal homogeneous spherical thin layer of neutral hydrogen gas with constant gas temperature $10^4 \text{K}$ and a homogeneous radial bulk velocity. In addition, the dust optical depth is set to $\tau_a  = (1 -  A_{\rm Ly{\alpha}}) \frac{Z}{Z_{\odot}} E_{\odot} N_{\rm H}$, where $E_{\odot} = 1.77 \times 10^{-21}\rm \text{cm}^{-2}$ is the ratio $\tau_a / N_{\rm H}$ for solar metallicity, $A_{\rm Ly_{\alpha}}$ = 0.39 is the albedo at the Ly$\alpha$ wavelength, $Z_{\odot} = 0.02$ \citep{Granato2000ApJ...542..710G}. For this, we vary the column density between log($N_{\text{H {\sc~i}}} /\rm \text{cm}^{-2}$)  between 17 and 22. The radiative transfer modelling of DLAs as an expanding shell results in an expansion velocity V$_{exp}$ = 30$_{-12}^{+55} \rm  km\ s^{-1}$, and a dust optical depth of log($\tau_{\alpha}$)= 0.20$_{- 0.17}^{+0.10}$. The posterior distributions of the fitted parameter are shown in Figure~\ref{zelda}.

The similar velocity offset, gas kinematics, and tight SFR-MS relation between the DLAs and LAEs hint at a close association between the two populations. Considering DLAs as LAEs, for a typical characteristic luminosity of LAEs, L$^{\star} = 5 \times 10^{42} \rm \text{erg}\ \text{s}^{-1}$ \citep{Cassata2011A&A...525A.143C}, our average Ly$\alpha$ emission imply that  DLAs traces the low-luminosity population, typically $\sim$0.02$L^{\star}$, of  LAEs. The long-slit spectroscopy efforts have detected the galaxy counterparts at small impact parameters of $< 15$kpc, suggesting a close association with the LAEs \citep{christensen_2014, krogager2017consensus}. Moreover, the MUSE surveys with a large 3D wide field of view have confirmed the association between DLAs and LAEs even at distances exceeding 50 kpc \citep{Fumagalli2017MNRAS.471.3686F,rubin2015dissecting,mackenzie2019linking}. In contrast, at $z \gtrsim 4 $  DLA galaxies are generally found at impact parameters of $\approx$15 - 60 kpc \citep{neeleman2017c,neeleman2019c,kaur2024hi}, suggesting their origin in CGM gas. Note that, the impact parameter is found to be decreasing with the increasing column density, thus, by selecting the log($N_{\text{H {\sc~i}}} / \rm \text{cm}^{-2}$)  $\ge 21$ we are more likely to probe the small impact parameters \citep[see][]{krogager2017consensus, Rahmati2014MNRAS.438..529R}. Moreover, the environment also plays a crucial role, as observation suggests higher detection of LAEs in environments with high column density absorbers, indicating that LAEs may trace larger-scale structures that include DLAs \citep{Lofthouse2020MNRAS.491.2057L}.  The same picture is corroborated by the EAGLE cosmological simulations, which indicate that the DLAs trace neutral regions within halos with characteristic masses of \(\text{M}_h \approx 10^{11} - 10^{12} \text{M}_\odot\) and likely associated with the LAEs where both are influenced by the same underlying dark matter halo \citep{font2012large}.  A follow-up study to trace galaxy counterparts at small impact parameters for a metallicity-unbiased DLA sample is crucial to uncover the faint-end population of LAEs.

 \begin{acknowledgements}
LCH was supported by the National Science Foundation of China (11991052, 12233001), the National Key R\&D Program of China (2022YFF0503401), and the China Manned Space Project (CMS-CSST-2021-A04, CMS-CSST-2021-A06). HC and Dharmender are grateful to  IUCAA for the hospitality under IUCAA associate programme. 
 \end{acknowledgements}
\bibliographystyle{aa} 
\bibliography{aa} 
 \begin{appendix}
    \section{Uncorrected 3$\sigma$-clipped weighted mean profile and zELDA parameter distribution for Ly$\alpha$ emission} 
\begin{figure}
    \centering
\includegraphics[width=0.5\textwidth]{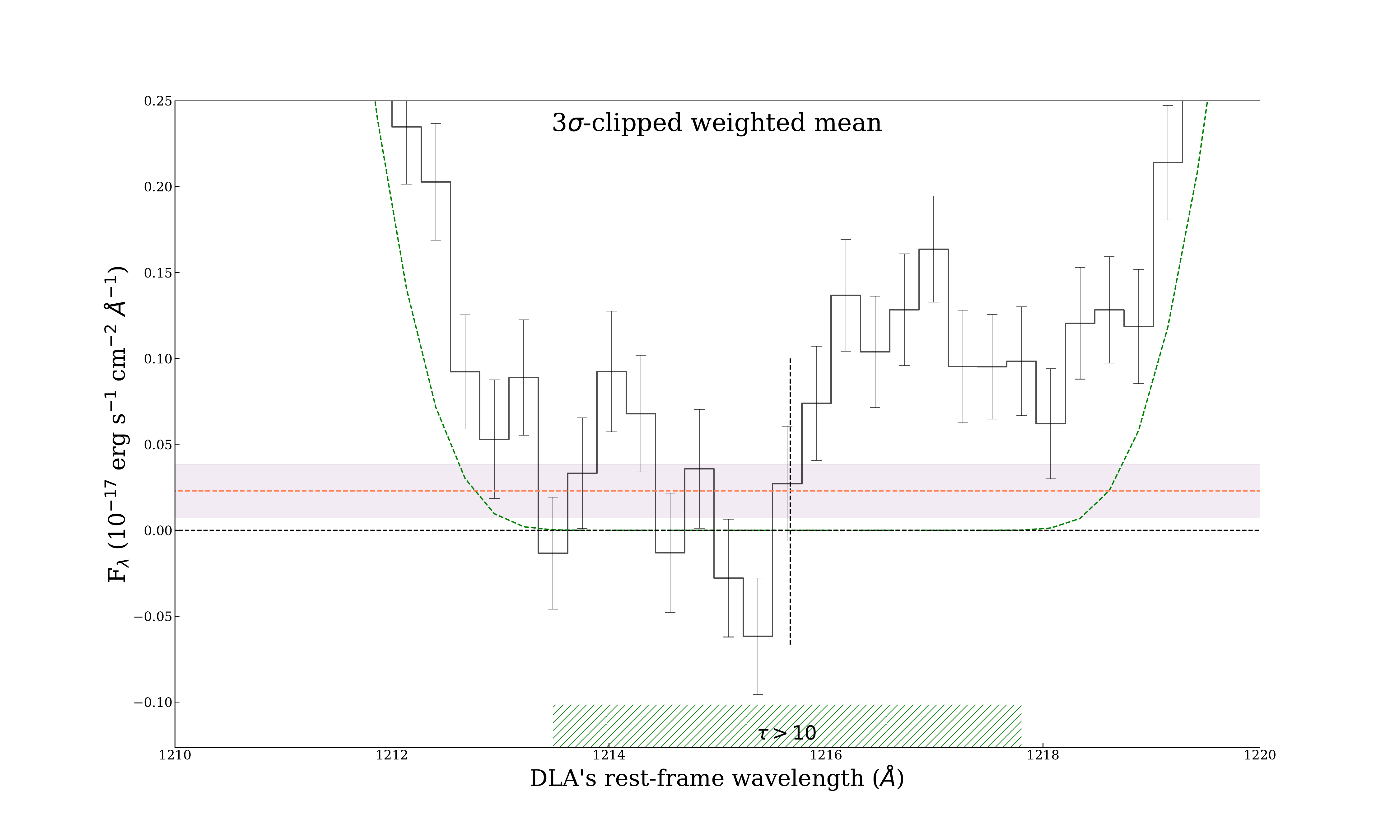}
    \caption{The 3$\sigma$-clipped weighted mean profile without zero-level correction. The average flux in the dark core of the DLA trough, blueward of the line center, is shown as {\it dashed} line, along with the mean error.} 
    \label{stack_nonzero}
\end{figure}
\begin{figure}
    \centering
    \includegraphics[width=0.47\textwidth]{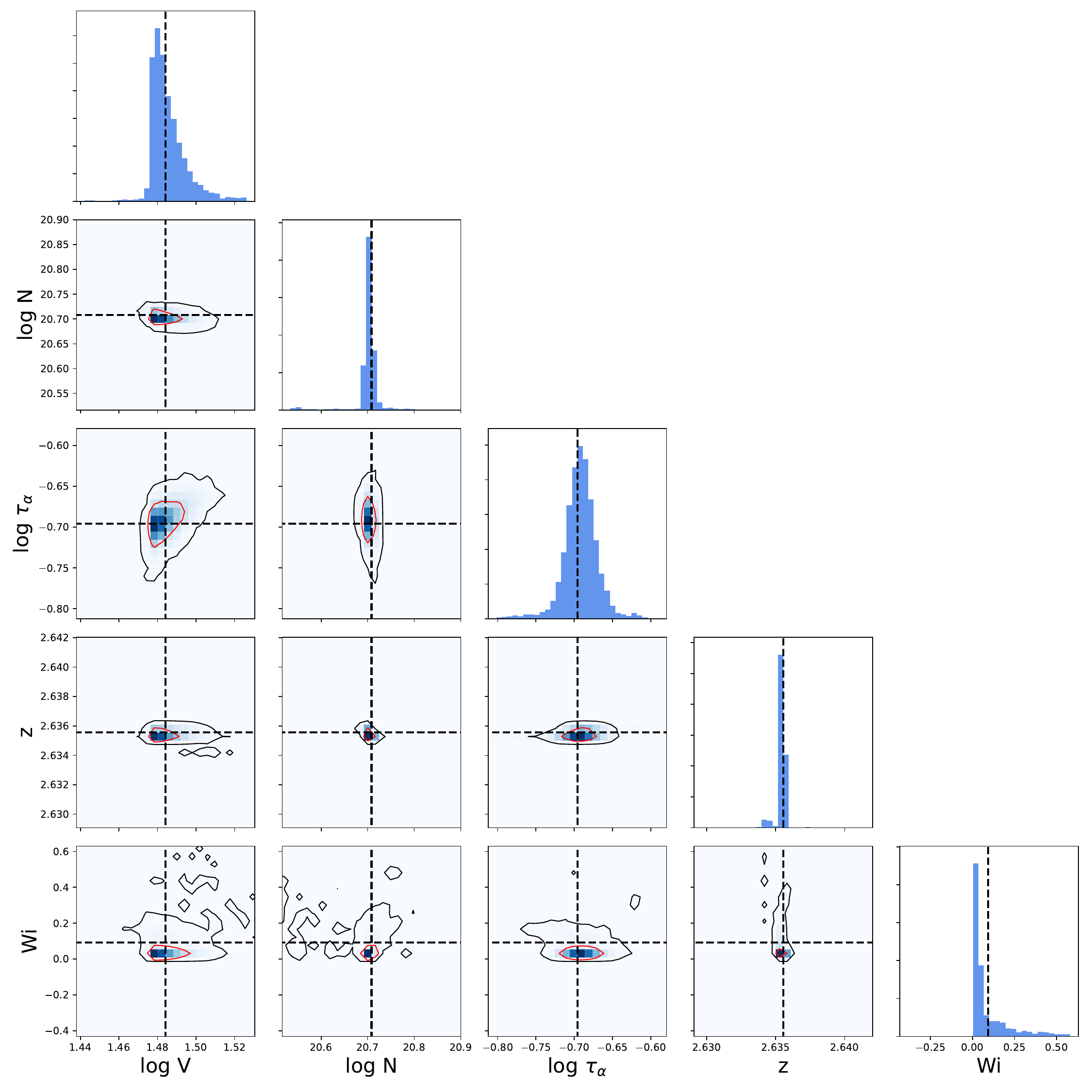}
    \caption{Posterior distributions for free parameters from zELDA radiative transfer modeling of Ly$\alpha$ emission line profile in 3$\sigma$-clipped weighted mean stack, at redshift 2.635, V$_{exp}$ = $30^{+55}_{-12} \text{km s}^{-1}$, log($N_{\text{H {\sc~i}}} /\rm \text{cm}^{-2}) = {20.7_{-0.01}^{+0.53}}$ , and $\tau_\alpha = 0.20^{+0.10}_{-0.17}$. The {\it grey} curves indicate the 1$\sigma$ and 2$\sigma$ contours of the 2D distributions.} 
    \label{zelda}
\end{figure}
 \end{appendix}
\end{document}